\documentclass[]{jfm}

\usepackage{graphicx}
\usepackage{bm}% bold math
\usepackage{newtxtext}
\usepackage{newtxmath}
\usepackage{natbib}
\usepackage{amsmath}
\usepackage{empheq}
\usepackage{hyperref}
\usepackage{subfigure}
\usepackage{dcolumn}% Align table columns on decimal point
\usepackage{physics}
\usepackage{tcolorbox}

\RequirePackage{color}
\hypersetup{
  colorlinks = true,
  urlcolor   = blue,
  citecolor  = black,
}

\newcommand{\RomanNumeralCaps}[1]
\linenumbers

\title{The Role of Vortex Stretching in Drag Reduction of Polymer-Laden Turbulent Flow}

\author{
   Wouter J. T. Bos\aff{1},
  \corresp{\email{wouter.bos@ec-lyon.fr}}
   Xuan Shao\aff{2,3},
   Tong Wu\aff{4},
and   Le Fang\aff{2,3}
  \corresp{\email{le.fang@buaa.edu.cn}}
}
\affiliation{
  \aff{1} CNRS, École Centrale de Lyon, INSA Lyon, Université Claude Bernard Lyon 1, LMFA, UMR5509, 69134 Écully, France
  \aff{2}Laboratory of Complex Systems, Ecole Centrale de P\'{e}kin, Beihang University, Beijing 100191, China
  \aff{3}Research Institute of Aero-Engine, Beihang University, Beijing 100191, China
  \aff{4}Theoretical Physics I, University of Bayreuth, Universitätsstr. 30, 95447 Bayreuth, Germany
 }

\begin{document}
\maketitle
\begin{abstract}
{\bf Abstract:} 
%In 1948, Toms showed that a small 
An addition of polymers can significantly reduce drag in wall-bounded turbulent flows, such as pipes or channels. This phenomenon is accompanied by a noticeable modification of the mean velocity profile. %No clear understanding exists allowing to predict this profile.  
Starting from the premise that  polymers reduce vortex-stretching, we derive a theoretical prediction for the mean-velocity profile. After assessing this prediction by  numerical experiments of turbulence with reduced vortex stretching, we show that the theory successfully describes experimental measurements of drag-reduction in pipe-flow.
\end{abstract}

%\begin{keywords}
%Turbulence, inhomogeneous,
%\end{keywords}
%
%{\bf MSC Codes }  {\it(Optional)} Please enter your MSC Codes here

% ==================================================
\section{Introduction}
% ==================================================

Turbulence rapidly drains energy from a fluid flow. Therefore, in many applications—such as aviation, naval and road transport, and industrial processes—a significant portion of supplied energy is lost to undesired and difficult-to-control turbulent motion. Minimizing turbulent drag is thus a key strategy for improving energetic efficiency. A spectacular example of turbulent drag-reduction is an effect first experimentally studied by B.A. Toms in the 1940s \citep{toms1949proceedings,toms1977early}. Adding a small amount of polymers to a turbulent pipe or channel flow can reduce the turbulent drag enormously, leading thereby to important savings of energy. 
%}.

Despite the large amount of investigations dedicated to the subject (see reviews of older contributions in  \citep{lumley1973drag,virk1975drag} and more recent work in \citep{procaccia2008colloquium,white2008mechanics,xi2019turbulent,dubief2023elasto}), the precise effect remains obscure. It is realized that the polymers interact with turbulent velocity fluctuations in a certain way which allows to reduce the overall energy dissipation and the momentum flux from the fluid towards the wall. Early efforts of De Gennes and Lumley pointed to possible mechanisms associated with the elastic properties of polymers \citep{de1986towards,tabor1986cascade} or the modification of the effective viscosity \citep{lumley1973drag} an idea further explored in for instance \cite{lvov2004drag,ryskin1987turbulent}. More recent studies have focused on spatio-temporal intermittency, or hibernating states associated with events of reduced drag-reduction \citep{xi2012dynamics,graham2004drag,whalley2017low}. Despite significant progress, none of these theories is entirely satisfactory or predictive.

%\section{Background}

%The subject has been extensively studied, and for comprehensive reviews of earlier contributions, we refer to the works of Lumley and Virk \cite{lumley1973drag,virk1975drag}. For more recent advancements, see \cite{procaccia2008colloquium,white2008mechanics,xi2019turbulent,dubief2023elasto}. 

%Despite eight decades of research on the subject, 

Even though the precise mechanism behind turbulent drag-reduction by polymers is unknown, the continued research efforts during eigth decades have yielded a wealth of insights. Indeed, several key features of dilute polymer-containing flows near boundaries are now well established, and here we highlight some of these features.

Polymer-laden flows are visco-elastic, which adds to the difficulty of normal, viscous flows. Due to their elastic nature, the polymers need an extensional flow to get stretched. If the typical time-scale of the flow is too large, the polymers will relax to their equilibrium coiled state. The dimensionless number which compares the elastic timescale to the flow-timescale is called the Weissenberg number $Wi$. At low Weissenberg numbers, polymers have minimal influence on the flow. %A certain level of turbulent activity is required to stretch the polymers from their coiled equilibrium state. Consequently, the Weissenberg number, which compares the characteristic timescales of the polymer to the timescale associated with the flow, must be sufficiently large for this stretching to occur \cite{degennes1974coil,watanabe2010coil}.

As $Wi$ increases beyond the coil-stretch transition \citep{degennes1974coil,watanabe2010coil}, the drag reduction becomes more significant, reaching, for large enough values of $Wi$, an upper limit associated with a flow configuration known as the maximum drag reduction (MDR) state \citep{virk1970ultimate,virk1975drag}. Notably, there appears to be a functional relationship between $Wi$ and the extent of drag reduction in parallel shear flows \citep{owolabi2017turbulent}. For large values of the drag-reduction, and up to the MDR state, the mean velocity profile in the near-wall region transitions from the logarithmic profile—characteristic of turbulent flow in Newtonian fluids—to a steeper profile. Historically, this profile was approximated using an alternative logarithmic expression \citep{virk1975drag}. However, more precise measurements have shown that the profile deviates from a logarithmic form \citep{white2012re}, exhibiting a convex shape in log-linear representation \citep{ptasinski2001experiments,owolabi2017turbulent}.

%Over the decades, numerous theories have been proposed to explain drag reduction by polymers. Without attempting to be exhaustive, we highlight several key approaches: theories based on effective viscosity \citep{lumley1973drag,lvov2004drag,ryskin1987turbulent}; elastic theories that incorporate modified multi-scale dynamics, where the smallest active flow scales are damped by elastic feedback on the flow \citep{tabor1986cascade,de1986towards}; and, more recently, theories linking drag reduction to quiescent flow states \citep{xi2012dynamics,graham2004drag}, which are intermittently triggered by polymers but also observed sporadically in Newtonian turbulence \citep{whalley2017low}. Despite significant progress, none of these theories is entirely satisfactory or predictive.

In this study, we focus on the role of vortex stretching reduction in drag reduction. The importance of vortex stretching has been mentioned in numerous studies over the years. As early as the works of %Landahl and Gadd
 \cite{landahl1973drag,gadd1968effects}, vortex stretching suppression was identified as a significant factor. Subsequent studies \citep{yarin1997mechanism, sureshkumar1997direct} expanded on this idea, while experiments demonstrated that material-line stretching was reduced in polymer-laden flows \citep{liberzon2005turbulence}. Recent numerical simulations confirm that polymers attenuate vortex stretching \citep{ur2022effect}, and experiments have shown that this attenuation is central to the mechanism of polymer drag reduction \citep{warwaruk2024local}.

Polymers reduce vortex stretching due to two key effects. First, it has been shown that rodlike passive particles or fibers align with the vorticity vector in turbulent flow \citep{pumir2011orientation,ni2014alignment}, and this is expected to hold true for polymers as well. Second, the presence of polymers significantly increases the extensional viscosity \citep{hinch1977mechanical,metzner1970stress,lindner2003obtain}. From these two observations—the alignment of polymers with vorticity and the increased extensional viscosity—it follows that polymers attenuate vortex stretching. This attenuation reduces drag since vortex stretching is one specific part of the nonlinear term, responsible for generating drag \citep{li2019decomposition}.

The idea that drag reduction is linked to increased extensional viscosity is not new. Most theoretical studies acknowledge that this increase is a principal effect of polymers on visco-elastic flow. In purely extensional flows, this increase can be interpreted as an effective viscosity enhancement \citep{ryskin1987calculation}. When flows are not purely extensional, modifying the effective viscosity inhomogeneously provides a coarse model for the influence of polymers in wall-bounded flow, albeit without accounting for the difference between extensional and shear stresses. This approach underpins Lumley’s theory \citep{lumley1973drag}, which can be refined using local energy balances and dimensional arguments to estimate viscosity changes \citep{lvov2004drag,procaccia2008colloquium}. However, firstly, this does not explain observations in homogeneous shearflow \citep{robert2010polymer,benzi2018polymers} where the viscosity should be statistically homogeneous and, secondly, turbulence is far from purely extensional. Incorporating a separate treatment  of the two different stresses into the Navier-Stokes equations remains challenging. Only recently has progress been made on this for the case of two-dimensional flows \citep{oliveira2024shear}, see also \cite{poole2023inelastic}.

The role of elasticity in polymer laden turbulence is still a debated subject. At moderate and high Reynolds numbers it is the fully stretched polymers which are responsible for the drag-reduction \cite{serafini2022drag}, thereby suggesting that elastic effects are not necessary to explain drag-reduction beyond the coil-stretch transition. However, around the MDR state, in particular at low Reynolds numbers, elastic instabilities might play an important role in maintaining a marginally unstable turbulent state  \citep{choueiri2018exceeding,dubief2023elasto}. We will come back to this point in the conclusion section. For the moment, we will ignore the effects of elasticity and consider flows beyond the coil-stretch transition.

In the present investigation, we avoid the complexities associated with altering the system's viscosity or modifying the viscous stress tensor. Instead, we focus directly on the vortex-stretching mechanism at the level of the governing equations. 
This approach neither relies on purely phenomenological arguments nor involves detailed simulations of viscoelastic turbulence.  %By adopting this approach, we avoid the treatment of the intricate interactions between polymers and turbulent flow. 
The strength of our method is its ability to isolate a specific aspect of polymer-turbulence interaction for specific analysis. Such an approach necessarily omits certain features of the interaction (such as elastic instabilities \cite{samanta2013elasto}, the effect of the weight-distribution of the polymers \citep{brandfellner2024quantitative,serafini2025role}, etc.), but it will prove valuable since it allows us to derive an analytical prediction of the inertial velocity profile. We will then compare this prediction to specifically designed numerical experiments, before comparing it to laboratory experiments of polymer-laden turbulent pipeflow.

% and second, to numerically integrate the modified Navier-Stokes equations to validate this prediction.

%In the present article the underlying mechanism of drag-reduction will be elucidated. The main ideas which will lead to the understanding of the effects are the following. Firstly, even though it seems a logical starting point to consider the momentum balance, described by the Navier-Stokes equations, a more intuitive way to unravel the effect of polymers on turbulence comes from the consideration of the vorticity balance. We argue that in this balance, it is the vortex stretching mechanism which is tamed, while the vorticity advection is almost unaffected. Starting from this premise, we use an almost-one-century-old model for the turbulent flux to obtain an analytical prediction for the mean velocity profile in turbulent channel flow (see Fig.~\ref{Fig:Visu}) with reduced vortex stretching. 

\begin{figure*}
\centering
\subfigure[]{\includegraphics[width=0.5\columnwidth]{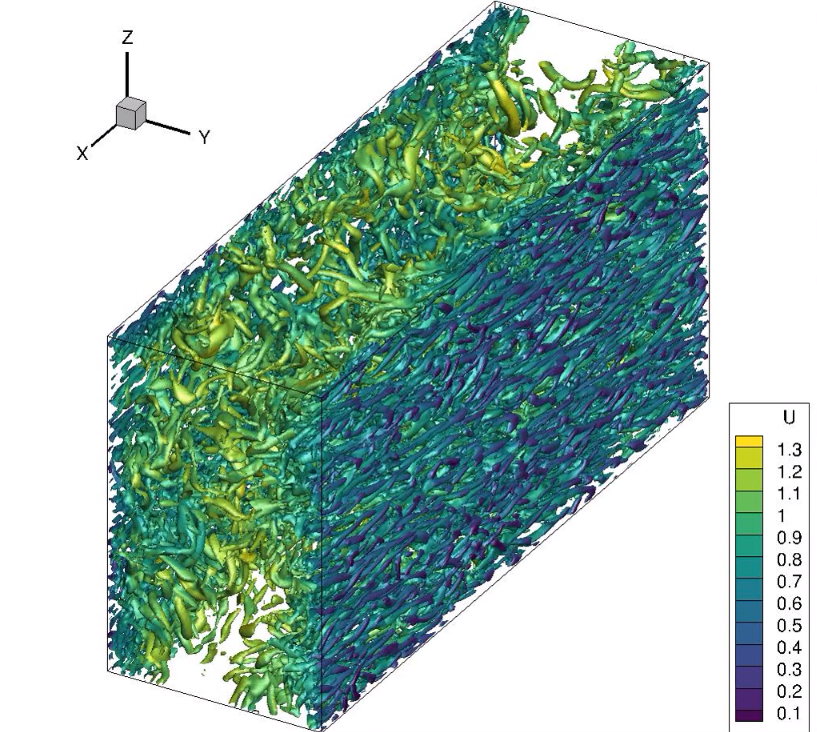}}~
\subfigure[]{\includegraphics[width=0.5\columnwidth]{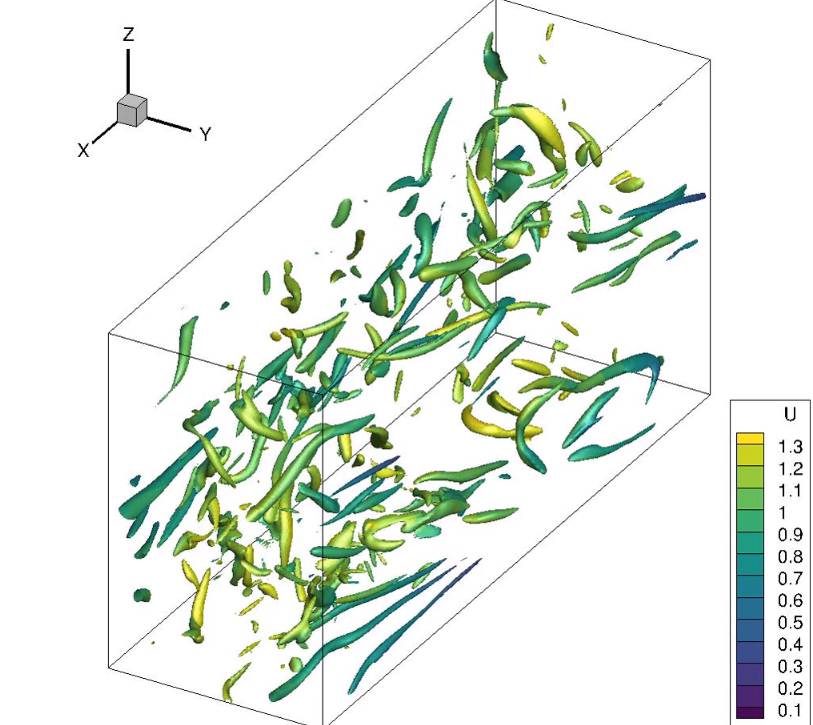}}
\caption{\label{Fig:Visu} Plane-Poiseuille flow driven by an imposed pressure gradient in the x-direction. %The steep gradients of the velocity due to the no-slip conditions at the wall lead to instabilities and turbulent motion. 
Vortical motion is amplified in three dimensional turbulence by vortex-stretching.  The flow visualizations show iso-surfaces of the $Q$-criterion  for $Q=1$ (see main text), colored by the local streamwise velocity. (a) Traditional channel flow. (b) Channel flow with reduced vortex stretching ($\lambda=0.35$ in Eq.~\eqref{eq:nslam}).}
\end{figure*}

\section{A law of the wall for turbulence with tamed vortex-stretching.}

\subsection{Derivation of the mean-profile}

To model the influence of reduced vortex stretching, we write the  evolution equation of the fluid vorticity  in the form,
\begin{equation}\label{eq:nslam}
\frac{\partial \bm \omega}{\partial t}+\bm u\cdot \bm \nabla \bm \omega=(1-\lambda)\bm \omega \cdot \bm \nabla \bm u -\lambda\nabla P_\omega+\nu \Delta \bm \omega,
\end{equation}
where the vorticity $\bm \omega$ is the curl of the velocity $\bm u$, $\nu$ the kinematic viscosity and $P_\omega$ a lagrange multiplier, ensuring $\nabla\cdot \bm \omega=0$  \citep{wu2023point}. A first property of this equation is that the left-hand-side, representing the advection of the vorticity, is not modified. Intuitively, this represent the fact that the advection of a small fluid volume should be largely independent (at leading order) of what molecular particles, particles or fibers the fluid particle contains. 
We hypothesize that this is different for the vortex-stretching term, since it is the term responsible for enstrophy generation, an effect dominated by extensional motion aligned with the vorticity. Indeed, the extensional stress is dramatically influenced by the presence of polymers in the flow. 

In the case $\lambda=0$ we retrieve the classical Navier-Stokes vorticity equation. The system \eqref{eq:nslam} for $\lambda=1$ was introduced in \citep{bos2021three,wu2021statistical} and studied in statistically homogeneous flow. In the present investigation we consider Eq.~\eqref{eq:nslam} in wall-bounded flow as the simplest model for polymer laden turbulence: we suppress vortex stretching partially or completely, by allowing values $0\leq \lambda\leq 1$. 

Before carrying out numerical experiments, we will first extend the classical arguments of Prandtl and von K\'arm\'an \citep{karman1930mechanische} to our system, in order to anticipate how near-wall scaling is modified by the attenuation of vortex stretching. For this we follow the same reasoning as von K\'arm\'an in his 1930 paper, which introduced the logarithmic law of the wall (or log-law). 

We focus on fully developed channel flow, with $x,y,z$ the streamwise, wall-normal and spanwise directions, respectively.  All average quantities are stationary and homogeneous in streamwise and spanwise direction. The mean velocity and vorticity fields are
\begin{equation}
\langle \bm u \rangle=U(y)\bm e_x~~~~~~~~\langle \bm \omega \rangle=-\frac{dU(y)}{dy} \bm e_z,
\end{equation}
respectively. From Eq.~\eqref{eq:nslam} we derive  the exact mean vorticity balance $\partial_t \langle \omega_z \rangle$,
\begin{eqnarray}\label{eq:momentumNVSl}
0=\lambda\frac{d\langle \omega_z'v' \rangle}{d y}+(\lambda-1)\frac{d^2\langle u'v' \rangle}{d y^2} +\nu \frac{d^3 U}{d y^3}. 
\end{eqnarray}
Neglecting viscous effects as usual in the derivation of the log-law, we proceed by modeling the turbulent fluxes of mean-vorticity and momentum, $\langle \omega_z'v' \rangle$ and $\langle u'v' \rangle$ using a classical mixing-length model,
\begin{eqnarray}
\langle a'v' \rangle=\nu_T\frac{\partial \langle a\rangle}{\partial y}, \textrm{ ~with~ } \nu_T=(\kappa y)^2 \left|\frac{\partial U}{\partial y}\right|,
 \end{eqnarray}
where $a$ can be both the velocity or vorticity. This corresponds to a mixing-length $l= \kappa y $ proportional to the distance to the wall \citep{karman1930mechanische}. In principle different values can be used for the constant in the mixing-length estimate for vorticity or velocity \citep{Hinze}, but this will not qualitatively change the outcome of this analysis. We obtain, since $d\langle \omega_z\rangle/dy=-d^2U/dy^2$,
\begin{eqnarray}\label{eq:lambdalaw0}
0=\kappa^2 \lambda y^2 \frac{d^2U}{dy^2}\frac{dU}{dy}+(1-\lambda)\kappa^2 \frac{d}{dy}\left(y^2\frac{dU}{dy}\frac{dU}{dy}\right).
\end{eqnarray}
Eq.~\eqref{eq:lambdalaw0} can be solved, yielding the solution
%\begin{eqnarray}\label{eq:lambdalaw}
%U^+(y^+)&=\begin{cases}
%\frac{1}{\kappa \Gamma}\left(\right)+U^+_0 ~~~&\textrm{for $\lambda\neq 0$} \label{eq:lambdalawa}\\
%\frac{1}{\kappa}\ln(y^+)+U^+_0 &\textrm{for $\lambda=0$},
%\end{cases}\\
\begin{eqnarray}\label{eq:lambdalaw}
U^+(y^+)=\frac{1}{\kappa }\frac{{y^+}^{\Gamma}-1}{\Gamma}+U^+_0
\end{eqnarray}
with 
\begin{equation}
\Gamma=\lambda/(2-\lambda)
\end{equation}
where quantities are expressed in wall-units ($U^+=U/u_\tau$, with $u_\tau=(\nabla \langle p\rangle h/\rho)^{1/2}$, $y^+=y u_\tau/\nu$). In the derivation, an integration constant is fixed, as customary, by the observation that in the inertial layer of newtonian channel flow turbulence $|\langle u'v'\rangle|\approx u_\tau^2$ \citep{Pope}. 

According to this phenomenological analysis we find thus power-law scaling with an exponent $\lambda/(2-\lambda)$  for $0<\lambda \le 1$. For $\lambda=0$ we find as a special case the logarithmic profile. Indeed, in the limit $\Gamma\downarrow 0$, we find that
\begin{eqnarray}\label{eq:lambdalawlog}
\lim_{\Gamma\rightarrow 0}U^+(y^+)=\frac{1}{\kappa }\ln(y^+)+U^+_0.
\end{eqnarray}  
%A remarkable property of  Eq.~\eqref{eq:lambdalaw} is that in our predictions the von-K\'arm\'an constant $\kappa$ is still the central constant which determines the velocity profile.  
The general $\lambda$-dependent mean velocity profile \eqref{eq:lambdalaw} is the main analytical prediction which we will compare first with direct numerical simulations of Eq.~\eqref{eq:nslam} and subsequently with experimental results from literature on polymer turbulence.

\begin{figure}
\centering
\subfigure[]{\includegraphics[width=0.5\columnwidth]{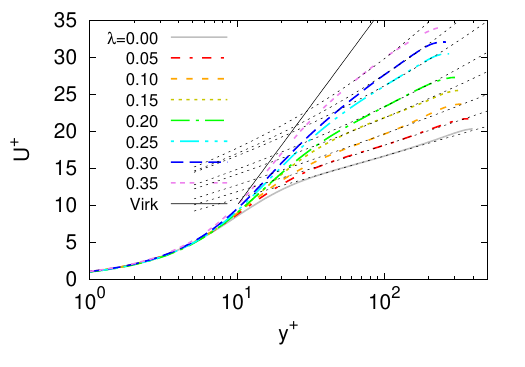}}~
\subfigure[]{\includegraphics[width=0.5\columnwidth]{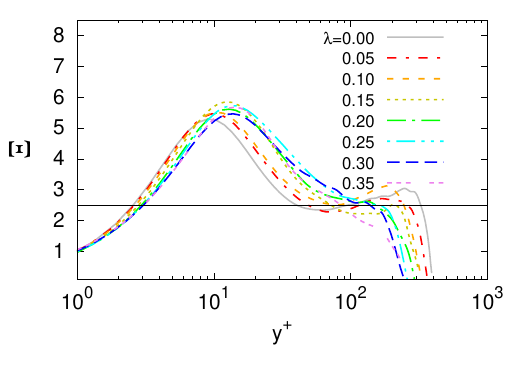}}
\caption{\label{Fig:Umean} Comparison of analytical predictions with our DNS results. (a) Mean velocity profiles in log-linear representation. The values of $U_0^+$ are for $\lambda\in [0,0.35]$: $5.1; 5.9; 6.6; 7.8; 8.3; 9.8; 9.9; 10.3$.
(b) The generalized indicator function  $\Xi=(y^+)^{1-\lambda/(2-\lambda)}dU^+(y^+)/dy^+$, Eq.~\eqref{eq:indicator} for the different velocity profiles in (a).}
\end{figure}

\subsection{Numerical experiments of turbulence with reduced vortex stretching} 

To assess our predictions, we carry out direct numerical simulations of Eq.~\eqref{eq:nslam} in plane channel flow. The walls are separated by a width $2h$, the width of the domain is $\pi h$ and the length $2\pi h$. Boundary conditions are no-slip at the wall and periodic in streamwise and spanwise directions. The code is based on a classical Fourier-Chebyshev formulation with a resolution of $(N_x=256,N_y=192,N_z=192)$ grid points. 
 Further details can be found in the %supplemental material. 
Appendix. Computations are carried out for $R_\tau=395$ at a constant mass flow rate. Simulations are run until a statistically steady state is obtained for the Navier-Stokes case $\lambda=0$, as measured by observing the wall shear-stress. Starting from this steady-state, $\lambda$ is varied in the range $\lambda \in [0,0.5]$ with steps of $\delta \lambda=0.05$. For simulations $\lambda \leq 0.35$ turbulent steady states are observed, while for $\lambda \geq 0.4$ flows relaminarized.

In Fig.~\ref{Fig:Visu} we show flow visualizations using the $Q$-criterion \citep{hunt1988eddies} for $Q\equiv \Delta p/2=1$. Comparing the Navier-Stokes case (a) with the case $\lambda=0.35$ (b), it is observed that in the second case the vortex intensity has reduced dramatically. Indeed, the vortex-stretching term is responsible for enstrophy production, and reducing it leads to the intuitive effect of less vortical activity.

In Fig.~\ref{Fig:Umean}(a) we show mean velocity profiles for the turbulent cases, determined by averaging during the statistically steady state. Also shown is our prediction, where $\kappa=0.4$ and the only adjustable parameter is the velocity offset $U^+_0(\lambda)$, which is fitted to obtain a best fit in the region where the generalized indicator function (see below) is in the vicinity of $1/\kappa$. Even though the Reynolds number is rather small, the predicted profiles describe the data well. The relation between the values $U^+_0(\lambda)$  and $\lambda$ is approximately linear, 
\begin{equation}\label{eq:U0}
U^+_0(\lambda)=U^+_0(0)+\lambda b_U,
\end{equation}
with $U^+_0(0)=5.2\pm 0.2$ and $b_U=15.8\pm 1$.

In order to assess our predictions independently from the virtual origin $U_0(\lambda)$, we introduce a generalized indicator function
\begin{equation}\label{eq:indicator}
\Xi[U,y]=y^{1-\Gamma}\frac{dU(y)}{dy}
\end{equation}
where $\Gamma=\lambda/(2-\lambda)$. For the value $\lambda=0$ this simplifies to the commonly used indicator function to verify the existence of a logarithmic profile (e.g. \citep{white2012re,laadhari2019refinement}). For $\lambda\neq 0$, this function should allow to identify the powerlaw behavior \eqref{eq:lambdalaw} by a plateau of value $1/\kappa$ independent from adjustable parameters other than the value $\lambda$, which is fixed for each simulation.

Even though the Reynolds number is too low to have a substantial inertial profile, 
Fig.~\ref{Fig:Umean}(b) shows that all 7 simulations coincide near the value $1/\kappa\approx 2.5$ around $y^+=100$, supporting the idea that $\kappa$ plays a central role, irrespective of the amount of vortex stretching that is suppressed in the governing equations. 

\section{Drag-reduction and vortex-stretching.}

\subsection{Similarities between the modified Navier-Stokes equations and visco-elastic turbulence}

The extended von Kármán phenomenology appears to capture the behavior of Eq.~\eqref{eq:nslam}, elucidating how reduced vortex stretching modifies the mean-velocity profile. The next question is: how relevant is Eq.~\eqref{eq:nslam} for modeling polymer turbulence?

Since Eq.~\eqref{eq:nslam} lacks derivation from a polymer stress constitutive relation (an important question for future research), its relevance must be inferred by comparing its properties to those of real flows. Before attempting such comparison, let us discuss certain similarities between polymer-laden turbulence and Eq.~\eqref{eq:nslam}.

First, the steepened velocity profile in our system qualitatively resembles profiles in polymer turbulence. %, especially for rod-like polymers. 
Although our approach does not model the coil-stretch transition \citep{degennes1974coil}, the continuous profile steepening aligns with high drag reduction regimes \citep{warholic1999influence}.

Second, our system stabilizes at a marginally stable state for $\lambda\approx 0.35$, with relaminarization at higher values. While polymer turbulence also reaches an asymptotic (MDR) state, the flow does not relaminarize at high Reynolds numbers. It is plausible that exceeding MDR suppresses vortex stretching to the point of relaminarization, at which polymers relax to their coiled equilibrium state. This would cause the flow to revert to Newtonian-like properties, reintroducing instabilities and turbulence. Local spatio-temporal relaminarization could correspond to hibernating turbulent states \citep{xi2012dynamics,graham2004drag}.

Third, drag reduction can occur for both elastic and rod-like polymers \citep{berman1978drag,paschkewitz2004numerical,japper2009turbulent}, suggesting that polymer-induced drag reduction is not purely elastic in origin. Our approach, lacking relaxation time scales (inherent to elastic effects) is consistent with this observation.

%reproduces an MDR-like state before relaminarization,

Fourth, simulations of viscoelastic isotropic turbulence and turbulence without vortex stretching reveal shared features. FENE-P model (finite extensible nonlinear elastic model with Peterlin's appoximation) simulations show that at high Weissenberg numbers, the kinetic energy spectrum transitions from a $k^{-5/3}$ scaling to $k^{-3}$ \citep{valente2014effect,valente2016energy}, as also observed in channel flows \citep{mitishita2023statistics}. This $k^{-3}$ scaling, linked to enstrophy conservation, is similarly observed in turbulence without vortex stretching \citep{wu2022cascades}. Furthermore, advanced simulations \citep{watanabe2013hybrid} demonstrate that polymers suppress enstrophy production, promoting a conservative enstrophy cascade at high Reynolds numbers.

These similarities motivate further scrutiny of the connection between polymer drag reduction and Eq.~\eqref{eq:nslam}. We therefore compare its predictions to experimental results.

\begin{figure}
\centering
\subfigure[]{\includegraphics[width=0.5\columnwidth]{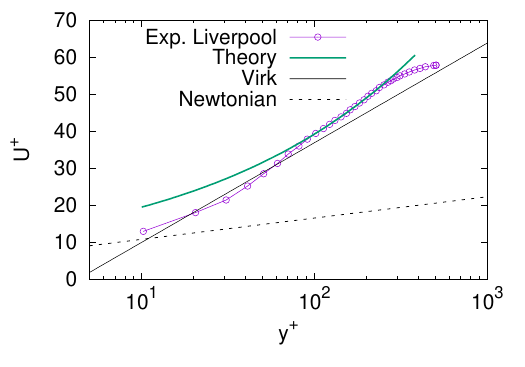}}~
\subfigure[]{\includegraphics[width=0.5\columnwidth]{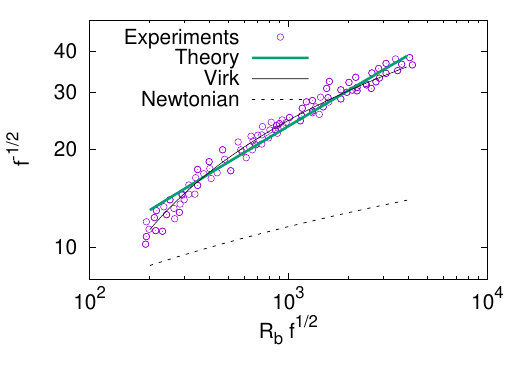}}
\caption{\label{Fig:Virk} Comparison of analytical predictions with Experimental results. (a) Mean velocity profile in pipe flow at MDR \citep{owolabi2017turbulent} compared to our powerlaw estimate. (b) Fanning friction factor in Prandtl-von Karman coordinates. Experimental results from smooth pipe measurements at MDR \citep{virk1970ultimate}.}
\end{figure}

\subsection{Comparison with experiments of visco-elastic shear flow.}
%pipe-flow.}

We compare our predictions to measures in pipe flow, which is, historically, the most investigated case.  We first compare to the velocity profile in pipe flow at MDR measured experimentally at the university of Liverpool \citep{owolabi2017turbulent} in Fig.~\ref{Fig:Virk}(a). Fitting our prediction Eq.~\eqref{eq:lambdalaw} in the range $100\le y^+\le 250$  we obtain $\Gamma\equiv\lambda/(2-\lambda)=0.349\pm 0.001$ and $U^+_0=3.5\pm 0.13$. Agreement in these scales is convincing and clearly superior to Virk's approximation.   

A larger amount of data is available on the friction-factor, which is more easily measured in experiments. We will here compare to the pioneering compilation of results by \cite{virk1970ultimate}. The friction-factor is determined by measuring the pressure drop $\delta p$ over a pipe of length $L$. Introducing the bulk velocity $U_b$ (velocity averaged over the pipe cross-section of radius $R$) the friction factor is
\begin{equation}
f=2\left(\frac{u_\tau}{U_b}\right)^2,
\end{equation}
where $u_\tau=\sqrt{R\delta p/(2\rho L)}$. Experiments are then performed to measure the value of $f$ as a function of the bulk Reynolds number $R_b=2R U_b/\nu$. Results are traditionally plotted, in Prandtl-von K\'arm\'an representation, $f^{-1/2}$ as a function of $R_b f^{1/2}(=2\sqrt{2}R_\tau)$.

We determine the bulk-velocity by integrating the velocity profile  over the pipe-cross-section, yielding after a change of variables $r=R-y$ and $y^+=((R-r)/R)R_\tau$ the relation
\begin{equation}
f^{-1/2}=\frac{\sqrt{2}}{ R_\tau}\int_0^{R_\tau}\left(1-\frac{y^+}{R_\tau}\right)U^+(y^+)dy^+.
\end{equation}
We now use expression \eqref{eq:lambdalaw} for $\lambda\neq 0$, to obtain 
\begin{equation}\label{eq:VirkGamma}
%f^{-1/2}=\frac{\sqrt{2}\left(\frac{R_b f^{1/2}}{2\sqrt{2}}\right)^\Gamma }%{\kappa \Gamma(\Gamma+1)(\Gamma+2)}+\frac{U^+_0}{ \sqrt{2}}.
%f^{-1/2}=\frac{2 \Lambda_\Gamma\left(R_b f^{1/2}/\sqrt{8}\right)-(\Gamma+3)}{\sqrt{2}\kappa (\Gamma+1)(\Gamma+2)}+\frac{U^+_0}{ \sqrt{2}}
f^{-1/2}=\frac{2 \left(\left[R_b f^{1/2}/\sqrt{8}\right]^{\Gamma}-1\right)/\Gamma-(\Gamma+3)}{\sqrt{2}\kappa (\Gamma+1)(\Gamma+2)}+\frac{U^+_0}{ \sqrt{2}}
\end{equation}
We show this expression in Fig.~\ref{Fig:Virk}(b). %Also shown is an empiric expression associated with Virk's logarithmic velocity profile. 
Fitting the data we find $\Gamma=0.348 \pm 0.003$ and $U^+_0=5.6\pm 0.4$. Also shown are the Newtonian expression $f^{-1/2}(x)=4 \log(x)-0.4$ and Virk's relation $f^{-1/2}(x)=19 \log(x)-32.4$. %Note that expression 

What we can say is that for large values of the Reynolds number, the agreement between  experiments of the friction factor and our theory is at least as good as using Virk's log-law, with the difference that our fit is based on a theoretical expression for the velocity profile (Eq.~\eqref{eq:lambdalaw}), valid for both Newtonian and visco-elastic flow, whereas Virk's expression is an empirical fit to the MDR data. For the velocity profile for the scales $100<y^+<250$ the agreement with our theory is clearly superior (Fig.~\ref{Fig:Virk}(a)). %Furthermore, Eq.~\eqref{eq:VirkGamma}  

%In particular the fact that the same value of $\Gamma$ is used in both the comparison of the friction factor Fig.~\ref{Fig:Virk}(a) and the velocity profile Fig.~\ref{Fig:Virk}(b) 

It is remarkable that we obtain, within error-bars, exactly the same value for $\Gamma$ in Fig.~\ref{Fig:Virk}(a) and Fig.~\ref{Fig:Virk}(b). This supports  the validity of Eq.~\eqref{eq:lambdalaw} to describe the mean profile at MDR.
The slightly different value of $U_0$ is not surprising since we obtain the estimate for $U_b$ by integrating over the full pipe cross-section although the power law does not hold for $y^+$ near the wall and the center. Another nice feature is that Eq.~\eqref{eq:VirkGamma} holds also for the Newtonian case, when the value $\Gamma=0$ is used.

%%%%%%%%%%%%%%%%%%%%%%%ù
%%%%%%%%%%%%%%%%%%%%%%%ù
%%%%%%%%%%%%%%%%%%%%%%%ù

\begin{figure}
\centering
\includegraphics[width=0.5\columnwidth]{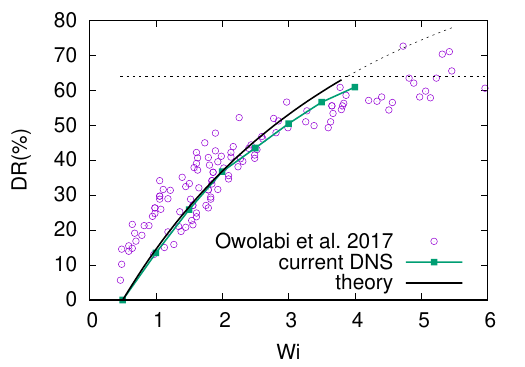}
\caption{\label{Fig:DR} Drag reduction as a function of the Weissenberg number. Comparison of experimental results of  \citep{owolabi2017turbulent} with our analytical predictions and numerical results for DNS with reduced vortex stretching. For the comparison we use a linear relation, $Wi-Wi_c=10\lambda$ between the reduction of vortex stretching $\lambda$ and $Wi$. The horizontal dashed line represents the MDR value ($DR=64\%$) corresponding to the fit to the experimental results in \cite{owolabi2017turbulent}.}
\end{figure}

A question we have not addressed is the link between the Weissenberg-number and the vortex-stretching parameter $\lambda$. At present, we have no direct physical relation between $\lambda$ and $Wi$ derived from a visco-elastic constitutive relation. 
In order to obtain a tentative link, we compare the results of our system with the compilation of \citep{owolabi2017turbulent}, who showed that the amount of drag-reduction in different pipes and channels could be linked by a close-to universal relation to $Wi$. The drag-reduction, for a fixed mass-flow rate, is defined as
\begin{equation}\label{eq:DR}
DR=\left(1-\left(\frac{u_\tau}{u_\tau^{(0)}}\right)^2 \right)
\end{equation}
where $u_\tau^{(0)}$ is the friction velocity for the reference case. 
In Fig.~\ref{Fig:DR} we show the experimental data-points for the interval $0<Wi<6$. We compare to the results of our DNS for different values of $\lambda$. To attempt a comparison and in absence of a physical model, we try to compare the data using the linear relation 
\begin{equation}
Wi-Wi_{c}=a \lambda,
\end{equation}
which is the simplest possible non-trivial relation, taking into account that the suppression of vortex-stretching should increase with the Weissenberg number, at least for values between the coil-stretch transition (at the critical Weissenberg number $Wi_c$) and MDR. We use the value $Wi_c=0.5$ and used the value $a=10$ which seems to give a good qualitative agreement.

To compare to the theory, we integrate the mean-velocity profile \eqref{eq:lambdalaw} over half the channel to get an estimate of the bulk-velocity
\begin{eqnarray}
U_b&=&\frac{u_\tau}{h}\int_0^h U(y) dy\nonumber\\ 
   &=&\frac{u_\tau}{\kappa(\Gamma +1)}\left[\frac{\left(\frac{hu_\tau}{\nu}\right)^{\Gamma}-1}{\Gamma}-1\right]+u_\tau U_0(\lambda)
\end{eqnarray}
yielding an implicit equation for $u_\tau$ for fixed $U_b$, which can be numerically solved, using Eq.~\eqref{eq:U0}. The resulting values of $u_\tau$ are used to compute the drag-reduction, shown in Fig.~\ref{Fig:DR}. The tendency is the same as the DNS and the experiments. Quantitative disagreement with the DNS can be attributed to the assumption to use Eq.~\eqref{eq:lambdalaw} over the complete channel, ignoring deviations near the wall and center and the domain. Using this relation between $\lambda$ and $Wi$, we find that the value $\Gamma = 0.349$ we obtained in Fig.~\ref{Fig:Virk} corresponds to $Wi_{MDR}\approx 5.6$, which is not inconsistent with the results in \cite{owolabi2017turbulent}.

\section{Conclusion and future issues} 

The complexity of polymer-laden turbulence is rooted in the multi-scale interaction of turbulence with polymer-molecules \citep{koide2024polymer} combined with the presence of walls, introducing both scale and position dependence. While advanced numerical experiments now provide access to detailed flow quantities in realistic geometries \citep{serafini2022drag}, progress in understanding requires drastic simplifications of flow physics to isolate specific features—an approach we have adopted. %Indeed, we implicitly assume that extensional damping primarily affects the vortex-stretching term. 

Introducing the new system of governing equations \eqref{eq:nslam} for the investigation of drag-reduction by polymers opens several avenues for exploration. A first and important one is the assessment of the assumption that strong extensional viscosity primarily acts to damp the term $\bm \omega \cdot \nabla \bm u$ in the vorticity dynamics. Such an assessment could start from the framework introduced in \cite{oliveira2024shear}. Further comparisons should investigate the implications of Eq.~\eqref{eq:nslam} for the detailed inhomogeneous and anistropic turbulence statistics \citep{escudier2009turbulent}.  

Perhaps even more intriguingly, investigating the self-sustaining mechanism of wall turbulence \citep{de2002dns,waleffe1997self} and the linear and nonlinear instability properties of shear flows with reduced vortex stretching, could provide theoretical insights into the nature of the MDR state.  While at low Reynolds numbers certain results indicate that elastic instabilities might determine the precise asymptotic value of MDR (\cite{dubief2023elasto} and references therein), this question is not settled at high Reynolds numbers. The MDR asymptote at higher Reynolds numbers might be determined by a subcritical transition \citep{morozov2005subcritical,morozov2007introductory}, and as such, its precise value is certainly not easy or impossible to determine by simple analytical reasoning. 

In this context, an interesting observation in our study is that we observe laminarization at relatively large value of the Reynolds number for sufficiently high values of $\lambda$, in the absence of elastic effects. Let us imagine what elastic effects would add in such a relaminarized state in an actual visco-elastic fluid. There polymers would relax back to their equilibrium state, which would have two direct influences: first, the relaxation could reintroduce kinetic energy into the flow when the polymer stress changes sign, as observed in isotropic visco-elastic turbulence \citep{nguyen2016small,valente2014effect} at small scales and high Weissenberg number, or in low-Reynolds shear-flows due to elastic instabilities \cite{choueiri2018exceeding}. This energy flux could
possibly perturb the relaminarized state. Second, after relaxation, the extensional viscosity would recover its newtonian value and the flow would become unstable again. Both effects would contribute to sustain a marginally unstable state, which would act as a visco-elastic fluid at the MDR asymptote.

These arguments support the ideas that firstly, (extensionally) viscous effects are the main actor in reaching the MDR state. Secondly, that elastic effects are essential to maintain a turbulent state once the MDR asymptote is reached. A more sophisticated model than Eq.~\eqref{eq:nslam} is needed to assess this reasoning and the development of such a model is left for further research.

\backsection[Ackowledgements]{We gratefully acknowledge discussions with Rob Poole, Faouzi Laadhari, Benjamin Miquel, Jinhan Xie and Chunxiao Xu. This work was supported by the National Natural Science Foundation of China (project approval nos. 12372214 and U2341231). The data that support the findings of this article are openly available.\\

For the purpose of Open Access, a CC-BY public copyright licence has been applied by the authors to the present document and will be applied to all subsequent versions up to the Author Accepted Manuscript arising from this submission.
}

\backsection[Declaration of interests]{The authors report no conflict of interest.}

\appendix
\section{Numerical Method}
% in Fig. \ref{fig:Sketch}. Instantaneous velocities in the streamwise ($x$), wall-normal ($y$), and spanwise ($z$) directions are denoted by $u$, $v$, and $w$, respectively, with their fluctuations represented by $u'$, $v'$, and $w'$. Periodic boundary conditions in the streamwise and spanwise directions ensure uniformity in these directions for fully developed turbulent Couette flow. No-slip and no-penetration boundary conditions are applied at the walls.

We carry out simulations of plane Poiseuille flow. The half-width of the channel is denoted $h$, the computational domain has dimensions of $L_x \times L_y \times L_z = 2\pi h \times 2h \times \pi h$ and employs $N_x \times N_y \times N_z = 256 \times 192 \times 192$ grid points. Uniform grids are applied in the streamwise and spanwise directions, while a non-uniform grid is used in the wall-normal (y) direction, defined by $y_j = \cos(\pi j / N_y)$ for $j = 0, 1, \ldots, N_y$. Note that in the numerical method $y=0$ is the centerline of the channel. The initial mean velocity profile is set as $U(y) =(3F/4)(1-y^2)$. By maintaining a constant mass flow rate $F=2$, the flow achieves a Reynolds number of $R_\tau = hu_\tau/\nu=395 $.  %with $U_c$ the mean centerline velocity 
with $\nu$ the kinematic viscosity and $u_\tau = \sqrt{\tau_w/\rho}$ the friction velocity. Here $\tau_w$ represents wall shear stress and $\rho$ denotes density. %The moving wall maintains a constant velocity $U_w$.

We conduct simulations with different values of $\lambda$, starting from a fully developed flow of conventional Navier-Stokes turbulence ($\lambda=0$). The numerical method for conventional turbulence  has been validated in previous studies \citep{Xu1996,Fang2011JoT,Chen2024PCFD}. Spatial discretization is conducted using the Fourier-Galerkin and Chebyshev-Gauss-Lobatto collocation methods: streamwise and spanwise directions are expanded using Fourier series, while the wall-normal direction utilizes Lagrange interpolation polynomials at Chebyshev-Gauss-Lobatto collocation points.

Before explaining the time advancement, let us start by formulating the evolution-equation for the velocity. Eq.~(1) of the manuscript
%\eqref{eq:nslam} 
is formulated for the evolution of the vorticity, whereas the numerical method we use is based on the time-advancement of the velocity. We therefore use the Biot-Savart operator to write
\begin{equation}\label{eq:BS}
 \frac{\partial \bm{u}}{\partial t}=\Delta^{-1}\nabla\times  \left(\frac{\partial \bm{\omega}}{\partial t}\right).
\end{equation}
We now write two evolution equations. The first equation corresponds to Navier-Stokes turbulence,
\begin{equation}\label{Eq.:Lamb}
  \left\{
    \begin{aligned}
      \frac{\partial \bm{u}}{\partial t} &= \bm{u} \times \bm{\omega}- \nabla \Pi + \nu\nabla^2 \bm{u} \\
      \Pi &= \frac{\bm{u} \cdot \bm{u}}{2} + \frac{p}{\rho}
    \end{aligned}.
  \right.
\end{equation}
 The second one corresponds to turbulence without vortex stretching,
\begin{equation}\label{Eq.:LambNVS}
  \left\{
    \begin{aligned}
      \frac{\partial \bm{u}}{\partial t}&=\Delta^{-1}\nabla\times((\bm{u} \cdot\nabla)\bm{\omega})-\nabla\Pi+ \nu\nabla^2 \bm{u} \\
      \Pi &= \frac{\bm{u} \cdot \bm{u}}{2} + \frac{p}{\rho}
    \end{aligned}.
  \right.
\end{equation}
where we used Eq.~\eqref{eq:BS}. Adding now the first and the second equation in proportions $\lambda$ and $1-\lambda$, we obtain the evolution of the velocity with partially removed vortex stretching,
\begin{equation}\label{Eq.:LambNVSPrec}
  \left\{
    \begin{aligned}
      \frac{\partial \bm{u}}{\partial t}&=\lambda\left(\Delta^{-1}\nabla\times((\bm{u} \cdot\nabla)\bm{\omega})\right)+(1-\lambda)(\bm{u}\times\bm{\omega})\nonumber \\
      &-\nabla\Pi+ \nu\nabla^2 \bm{u} \\
      \Pi &= \frac{\bm{u} \cdot \bm{u}}{2} + \frac{p}{\rho}
    \end{aligned}.
  \right.
\end{equation}
Now that we have a velocity-evolution-equation we will describe the time advancement in detail. 

The time-advancement scheme is based on a semi-implicit, three-step backward differentiation formula (that is, $J_e=3$) \citep{moin1978large,fan2022principles}.  During each time step, a third-order time-splitting method is employed, including three substeps that sequentially process the convective, pressure, and diffusion terms respectively. In the following parts, superscripts $s$, $s+\frac{1}{3}$, $s+\frac{2}{3}$, and $s+1$ will be used to denote the advancement from time step $s$ to $s+1$, corresponding to the substeps respectively.

In the present work without vortex stretching, the convective term in Eq. (\ref{Eq.:LambNVSPrec}) is advanced from $s$ to $s+\frac{1}{3}$ using two microsteps, by further introducing an instant $s+\frac{1}{6}$. The first microstep indicates the influence of the term $\lambda\left(\Delta^{-1}\nabla\times((\bm{u} \cdot\nabla)\bm{\omega})\right)$, written as
\begin{equation}\label{Eq.:ConTermNVS1}
  \frac{\Delta\bm{u}^{s+\frac{1}{6}}-\sum_{q=0}^{J_e-1} \alpha_q \Delta\bm{u}^{s-q}}{\mathrm{d}t}=\sum_{q=0}^{J_e-1} \gamma_q \left[\lambda(\nabla\times(\bm{u} \cdot\nabla)\bm{\omega})\right]^{s-q},
  \end{equation}
with no-slip boundary conditions
\begin{equation}\label{Eq.:ConNVSBoundary1}
    \left\{
      \begin{aligned}
%  &u_{y=1}=2\gamma_4\\
%  &u_{y=1}=\\
  &u_{y=\pm 1}=v_{y=\pm 1}=w_{y=\pm 1}=0
  \end{aligned}.
    \right.
\end{equation}
The second microstep indicates the influence of the term $(1-\lambda)(\bm{u}\times\bm{\omega})$
\begin{equation}\label{Eq.:ConTermNVS2}
  \frac{\bm{u}^{s+\frac{1}{3}}-\bm{u}^{s+\frac{1}{6}}}{\mathrm{d}t}=\sum_{q=0}^{J_e-1} \gamma_q \left[(1-\lambda)(\bm{u} \times \bm{\omega})\right]^{s-q},
  \end{equation}
with additional boundary conditions
\begin{equation}\label{Eq.:ConTermBoundary2}
    \frac{\bm{u}^{s+\frac{1}{3}}-\sum_{q=0}^{J_e-1} \alpha_q \bm{u}^{s-q}}{\mathrm{d}t}=\sum_{q=0}^{J_e-1} \gamma_q \left[(1-\lambda)(\bm{u} \times \bm{\omega})\right]^{s-q}.
\end{equation}

The calculation of the pressure term in Eq. (\ref{Eq.:LambNVSPrec}) (from $s+\frac{1}{3}$ to $s+\frac{2}{3}$) advances as
\begin{equation}\label{Eq.:PreTerm}
    \left\{
      \begin{aligned}
        \frac{\bm{u}^{s+\frac{2}{3}}-\bm{u}^{s+\frac{1}{3}}}{\mathrm{d}t}&=-\nabla \Pi^{s+\frac{2}{3}}\\
        \nabla \cdot \bm{u}^{s+\frac{2}{3}}&=0
      \end{aligned}.
    \right.
\end{equation}
for the internal field, with the wall boundary condition for the pressure step modified as
\begin{eqnarray}
\label{Eq.:PreBoundaryNVS}
\left( \nabla \Pi^{s+\frac{2}{3}} \right) \cdot \bm{e_y}= \left(\sum_{q=0}^{J_e-1} \gamma_q \left[(1-\lambda)(\bm{u} \times \bm{\omega})\right]^{s-q} \right. \nonumber\\
\left.
+\nu\sum_{q=0}^{J_e-1} \gamma_q\left(-\nabla \times\left(\nabla \times \bm{u}^{s-q}\right)\right)\right) \cdot \bm{e_y}
\end{eqnarray}
where Eq. (\ref{Eq.:PreBoundaryNVS}) involve the unit normal vector $\bm{e_y}$. The coefficients in Eqs. (\ref{Eq.:ConTermNVS1}), (\ref{Eq.:ConTermNVS2}), (\ref{Eq.:ConNVSBoundary1}), (\ref{Eq.:ConTermBoundary2}) and (\ref{Eq.:PreBoundaryNVS}) remain consistent with conventional Navier-Stokes turbulence in the case $\lambda=0$.

The viscous term, advancing from time step $s+\frac{2}{3}$ to $s+1$, is
\begin{equation}\label{Eq.:VisTerm}
  \frac{\gamma_3 \bm{u}^{s+1}-\bm{u}^{s+\frac{2}{3}}}{\mathrm{d}t}=\nu \nabla^2 \bm{u}^{s+1},
\end{equation} 
%with no-slip boundary condition for the velocity:
%\begin{equation}\label{Eq.:VisBoundary}
%  \left\{
%    \begin{aligned}
%&u_{y=1}=2\\
%&u_{y=-1}=v_{y=\pm 1}=w_{y=\pm 1}=0
%\end{aligned}.
%  \right.
%\end{equation}
The coefficients in the semi-implicit scheme $\alpha_0, \alpha_1, \alpha_2, \gamma_0, \gamma_1, \gamma_2 $ and  $\gamma_3$ are reported in \cite{karniadakis1991high}.

%\lambda_1, \lambda_2$ 
%and $\gamma_4$ are reported in Ref. \cite{moin1978large}.

%\bibliographystyle{unsrt}
%\bibliography{/home/wbos/PUBLI/biblio}

%\bibliographystyle{jfm}
%\bibliography{/home/wbos/PUBLI/biblio}

\end{document}